\documentclass{PoS}
\title{The full Schwinger-Dyson tower for random tensor models}

\newcommand{\logos}[4]{\raisebox{-.#4\height}{\includegraphics[height=#3 ex]{gfx_S/3/Logo#1_#2.pdf}}} 
 
\newcommand{\suml}{\sum\limits}

\newcommand{\logoB}[4]{\raisebox{-.#4\height}{\includegraphics[height=#3 ex]{gfx_S/3B/Logo#1_#2.pdf}}} 
  \newcommand{\iconoB}[4]{\raisebox{-.#4\height}{\includegraphics[height=#3 ex]{gfx_S/3B/Item#1_#2.pdf}}} 

\newcommand{\vch}{\raisebox{-.322\height}{\includegraphics[height=2.1ex]{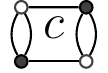}}}
\newcommand{\vbh}{\raisebox{-.322\height}{\includegraphics[height=2.1ex]{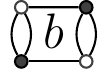}}}
\usepackage{braket}
\usepackage{mathtools}  
\usepackage{amssymb} 
\usepackage{bbold}
\usepackage{amsthm}
\usepackage{amsfonts}  	
\usepackage{stmaryrd}

\newcommand{\GDmelon}{G\hp{ 2}_{\raisebox{-.33\height}{\includegraphics[height=2.2ex]{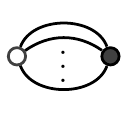}}}}

\newcommand{\vc}{\raisebox{-.322\height}{\includegraphics[height=2.1ex]{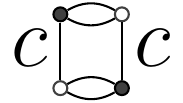}}}
\newcommand{\va}{\raisebox{-.322\height}{\includegraphics[height=2.3ex]{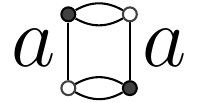}}}
\newcommand{\vuno}{\raisebox{-.322\height}{\includegraphics[height=2.3ex]{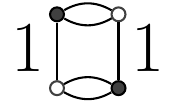}}}

\newcommand{\vcCT}{\raisebox{-.322\height}{\includegraphics[height=2.3ex]{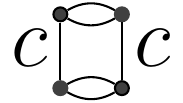}}}
  \newcommand{\vdos}{\raisebox{-.322\height}{\includegraphics[height=2.3ex]{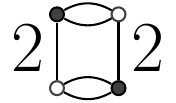}}}
  \newcommand{\vtres}{\raisebox{-.322\height}{\includegraphics[height=2.3ex]{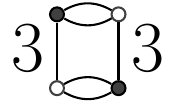}}}

\usepackage{xcolor}
\colorlet{texto}{black!50!gray} 

\def\tetrach{\tikz[baseline=.1ex,scale=.63]{
\fill[texto] (-.5ex,0) circle (1.5pt) coordinate (A);
\fill[texto]  (1ex,-1.5ex) circle (1.5pt) coordinate (B); 
\fill[texto]  (2.9ex,.8ex) circle (1.5pt) coordinate (C);
\fill[texto]  (1ex,3ex) circle (1.5pt) coordinate (D); 
\path 
    (A) edge (C)
      ;
      
\draw[draw=white,double=texto,double distance=.8 \pgflinewidth, thick] (B) to (D);
 \path (A) edge (B) 
      (B) edge  (C)
      (C) edge  (D) 
      (A) edge (D) 
      (B) edge (D)
      ;

\fill[texto] (1ex,3ex) circle (1pt) ;
\fill[texto]  (1ex,-1.5ex) circle (1pt) ;
\fill[texto]  (-.5ex,0) circle (1.5pt) coordinate (A);
\fill[texto] (1ex,-1.5ex) circle (1.5pt) coordinate (B); 
\fill[texto] (2.9ex,.8ex) circle (1.5pt) coordinate (C);
\fill[texto] (1ex,3ex) circle (1.5pt) coordinate (D); 
}%
}

\usepackage{amsmath}
\newtheorem{thm}{Theorem}[section]

\theoremstyle{definition}

\newtheorem{example}[thm]{Example}

\newcommand{\Vuno}{\raisebox{-.322\height}{\includegraphics[height=3.54ex]{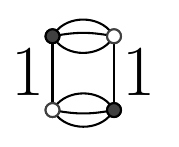}}}
\newcommand{\Vdos}{\raisebox{-.322\height}{\includegraphics[height=3.54ex]{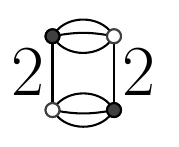}}}
\newcommand{\Vtres}{\raisebox{-.322\height}{\includegraphics[height=3.54ex]{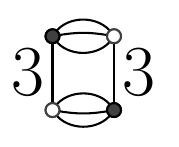}}}
\newcommand{\Vcuatro}{\raisebox{-.322\height}{\includegraphics[height=3.54ex]{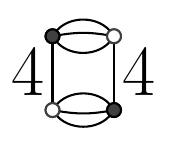}}}

  \newcommand{\logito}[4]{\raisebox{-.#4\height}{\includegraphics[height=#3 ex]{gfx_S/3/Logo#1_#2.pdf}}} 
\newcommand{\icono}[4]{\raisebox{-.#4\height}{\includegraphics[height=#3 ex]{gfx_S/3/Item#1_#2.pdf}}}

\newcommand{\meloncitito}{\icono{2}{Melon}{1.5}{20}}
\newcommand{\meloncito}{\icono{2}{Melon}{2}{24}}
\newcommand{\melon}{\logito{2}{Melon}{2.6}{3}}
\newcommand{\J}{ \hspace{2pt} \mbox{\textit{\textbf{j}}}}

\usepackage{amsbsy}

 \newcommand{\Sint}{S_{\mtr{int}}}
 \newcommand{\kthree}{\raisebox{-.2\height}{\includegraphics[width=2ex]{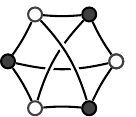}}}
 
 \newcommand{\kthreegg}{\raisebox{-.2\height}{\includegraphics[width=2.83ex]{graphs/3/Item6_K33.pdf}}}

 \newcommand{\kthreea}{\raisebox{-.12\height}{\includegraphics[width=2ex]{graphs/3/Item6_K33.pdf}}}
\newcommand{\fder}[2]{\frac{\delta #1}{\delta #2}}
\newcommand{\jj}{[J,\bJ\,]}

\newcommand{\xb}{\mathbf{x}}
\newcommand{\yb}{\mathbf{y}}
\newcommand{\zb}{\mathbf{z}}
\newcommand{\phat}{\mathbf{p}_{\hat a}}

\newcommand{\Xb}{\mathbf{X}}

\renewcommand{\sb}{\mathbf{s}}

 \usepackage{multicol,array}
\makeatletter
\def\moverlay{\mathpalette\mov@rlay}
\def\mov@rlay#1#2{\leavevmode\vtop{%
   \baselineskip\z@skip \lineskiplimit-\maxdimen
   \ialign{\hfil$\m@th#1##$\hfil\cr#2\crcr}}}
\newcommand{\charfusion}[3][\mathord]{
    #1{\ifx#1\mathop\vphantom{#2}\fi
        \mathpalette\mov@rlay{#2\cr#3}
      }
    \ifx#1\mathop\expandafter\displaylimits\fi}
\makeatother

\usepackage{rotating}

\usepackage{multirow}

\newcommand{\Aut}{\mathrm{Aut}}
\newcommand{\Autc}{\Aut_{\mathrm{c}}}
\renewcommand{\and}{\mbox{ and }}

\newcommand{\T}{\mathbb{T}}
\newcommand{\Tr}{\mathrm{Tr}}

\newcommand{\mtr}[1]{\mathrm{#1}}

\newcommand{\A}{\mathcal{A}}

\newcommand{\dervpar}[2]{\frac{\partial #1}{\partial #2}}

\newcommand{\G}{\mathcal{G}}

\newcommand{\B}{\mathcal{B}}
\newcommand{\C}{\mathbb{C}}

\newcommand{\inv}{^{-1}} 
\newcommand{\mtc}[1]{\mathcal{#1}} 
\newcommand{\Z}{\mathbb{Z}}

\newcommand{\Sym}{\mathrm{Sym}}
 
\newcommand{\hp}[1]{^{(#1)}}

\renewcommand{\phi}{\varphi}

\newcommand{\uni}[1]{\mathrm{U}(#1)}
\newcommand{\orth}[1]{\mathrm{O}(#1)}

 \usepackage{booktabs,array}
\usepackage{textcomp}
\usepackage{simplewick}
\usepackage{float}
\usepackage{xifthen}
\usepackage{sansmath}
\usepackage{marvosym}
\usepackage{tensor}
\usepackage{xparse}

\usepackage{amsfonts}
\usepackage{amssymb}

\usepackage{lastpage}
\usepackage{fancyhdr}
\usepackage{setspace}
\usepackage{tikz-cd}

\newcommand{\bJ}{\bar{J}}

\newcommand{\subircm}[1]{ \raisebox{.1cm}{#1\vphantom{A}}}

\ShortTitle{The full Schwinger-Dyson tower for random tensor models}

\author{\speaker{Carlos Ignacio P\'erez-S\'anchez} \thanks{The author's participation in the 
Corfu 2017 summer school and workshop was possible thanks to COST Action MP1405 Quantum Structure of Spacetime (QSpace).
The author wishes to thank the Corfu Summer Institute 2017 hospitality.
The author thanks George Zoupanos and for the invitation to write 
  this note. The Collaborative Research Center ``Groups, Geometry \& Actions'', SFB 878 (University of 
  M\"unster),
  is acknowledged for financial support. 
  Comments from and discussions with Romain Pascalie, Johannes Th\"urigen and Raimar Wulkenhaar 
  were fruitful for this review.
  }
  \\
  Mathematisches Institut der Westf\"alischen Wilhelms-Universit\"at,\\ Einsteinstra\ss e 62,
48149 M\"unster, Germany
  \\
  E-mail: \email{perezsan@uni-muenster.de}

  }

\abstract{
We treat random rank-$D$ tensor models as $D$-dimensional quantum field theories---tensor field theories (TFT)---and review some of their non-perturbative methods.  
We classify the correlation functions of complex tensor field theories by boundary graphs, sketch the derivation of the Ward-Takahashi identity and stress its relevance in the derivation of the tower of exact, analytic Schwinger-Dyson equations for all the correlation functions (with connected boundary) of TFTs with quartic pillow-like interactions.
}

\usepackage{tikz}
\FullConference{Corfu Summer Institute 2017 "School and Workshops on Elementary Particle Physics and Gravity"\\
  2-28 September 2017\\
  Corfu, Greece}

\begin{document}

\section{Introduction}

In ordinary Quantum Field Theory (QFT) the Schwinger-Dyson equations
account for the non-perturbative description of propagations and interactions,
expressed in terms of equations of motion for the Green's functions.  
Non-perturbative methods usually yield an infinite tower of coupled Schwinger-Dyson equations,
which is rarely solvable.  Some matrix models (or rather, matrix quantum
field theories) \cite{GW12} escape 
 this feature, though. 
The solvability of the real quartic matrix model heavily (but not exclusively) relies on 
the $N^{2-2g}$-expansion % $G\hp k=\sum_{g\in \Z_{\geq 0}} N^{2-2g} G\hp {k; g} $
of the Green's functions in the matrix size, $N$, which allows to derive a closed equation for 
the two-point function in the planar ($g=0$) sector and thereafter to determine 
the higher-point functions by algebraic recursions. 
The extension of these non-perturbative approach to other 
kind of theories that also possess an inverse-$N$
expansion is therefore intriguing, since there 
it is natural to test for solvability, at least in the 
large-$N$ limit. To such family belong \textit{(random) tensor models}.
\par

The matrix model description of 2D-quantum gravity \cite{dFGZ} inspired
tensor models \cite{Ambjorn} and random tensors \cite{guraubook}.
A colored structure on the tensors 
\cite{Gurau:2009tw} led to their
$1/N$-expansion. 
Beyond the random geometry and quantum gravity \cite{critical,GurauRyan, RTM_QG}
applications that tensor models had, 
the large-number-of-particles limit of the 
Sachdev-Ye-Kitaev (SYK) \cite{Kitaev} 
also unexpectedly received a  tensor model description 
\cite{WittenSYK,stanford,KlebanovTarnopolsky,Bonzom:2017pqs,BenGeloun:2017jbi}
\footnote{See \cite{NicoRivasseauCorfu} in these proceedings.}
and has become a tool in holography. \par 
This short article only describes non-perturbative 
QFT aspects of (complex) tensor models;
the reader is referred to the previous sources for 
a deeper physical approach.
\par

For a scalar theory with cubic and quartic interactions, for sake of concreteness, 
the Schwinger-Dyson equations (SDE) are recursions that describe the insertions 
of the $n$-point and $(n+1)$-point functions into the $(n-1)$-point function 
one has then terms of the form (see e.g. \cite{HuberSDE} and \cite[Fig. 1]{HuberNotes} )
The 1PI $2$-point function $\Gamma\hp 2$, for instance, satisfies:
\begin{align}
\raisebox{-.34\height}{
\includegraphics[width=.917\textwidth]{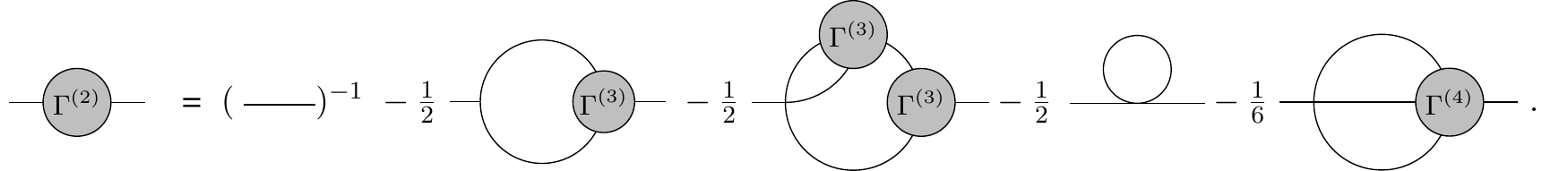}}
 \label{SDEscalar}
\end{align}

\begin{figure}\centering
\includegraphics[width=5.75cm]{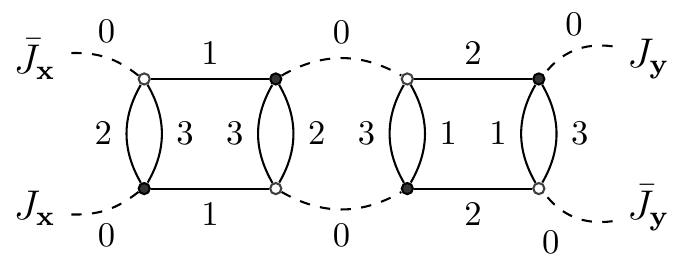}
 \caption{Connected 4-point $\phi_3^4$-Feynman graph with disconnected
 boundary. The dashed (or $0$ color) \label{fig:nia}
 transmits simultaneously the three colors. 
 Hence, momentum transmission is encoded by 
 $0a$-bicolored paths between external legs. 
 This draws a graph, the \textit{boundary graph}, which in this case is 
 disconnected. Thus, this graph contributes to the 
  $4$-point function $G\hp{4}_{\mtr m | \mtr m}$ or $G\hp{4}_{\mtr{disconn.}}$ (see also
   Table 1 for notation).
   }
\end{figure}

% \begin{align} \label{SDEscalar}
% % \bigg(
% \raisebox{-.374\height}
% {\tikz[scale=.77]{
% \draw[] (-1,0) --( 1,0 );
% \draw[black,fill=lightgray] (0,0) circle (.5)
% ;
% \node at (0,0) {\footnotesize $\Gamma\hp2$};
% }}  \,\,
%  & =\,\,
% \big(
% \raisebox{.0651cm}
% {\tikz[scale=.67]{
% \draw[] (-.61,.20) --( .6,.20 );
% % \fill[lightgray] (0,0) circle (.5)
% ;}}\big)^{-1} \,\,
% -\frac{1}{2}\,\,
% \raisebox{-.45\height}
% {\tikz[scale=.7	]{
%  \draw[] (-1.5,0) --(-1,0 );
%   \draw[] (2,0) --(1,0 );
%  \draw[] (0,0) circle (1);
% \draw[black,fill=lightgray] (1,0) circle (.5)
% ;
% \node at (1,0) {\footnotesize $\Gamma\hp3$};
% }}
% \,\,-\frac{1}{2}
% \,\,
% \raisebox{-.35\height}
% {\tikz[scale=.77]{
%  \draw[] (-1.5,0) --(-1,0 )
%  ;
%  \draw   (-1,0) arc (-90:0:1.1) ; 
%   \draw[] (2,0) --(1,0 );
%  \draw[] (0,0) circle (1);
% \draw[black,fill=lightgray] (1,0) circle (.5);
% \node at (1,0) {\footnotesize $\Gamma\hp3$};
% \draw[black,fill=lightgray] (0,1) circle (.5);
% \node at (0,1) {\footnotesize $\Gamma\hp3$};
% }} 
% \\ & \,\,\,\,\,\,\,\,\,\,\,\,
% -\frac{1}{2}\,\,\,
% \raisebox{.1035\height}
% {\tikz[scale=.77]{
% %  \draw[] (-1.5,0) --(-1,0 )
%  ;
%  \draw   (0,0) arc (-90:269.9:.5) ; 
%   \draw[] (-1,0) --(1.,0 );
% }}
% -\frac{1}{6}\,\,
% \raisebox{-.435\height}
% {\tikz[scale=.77]{
% %  \draw[] (-1.5,0) --(-1,0 )
%  ;
% %  \draw   (-1,0) arc (-90:0:1) ; 
%   \draw[] (-1.5,0) --(2,0 );
%  \draw[] (0,0) circle (1);
% \draw[black,fill=lightgray] (1,0) circle (.5);
% \node at (1,0) {\footnotesize $\Gamma\hp4$};
% }}\,\,.  \,\,\,\,\, \nonumber
% \end{align}
Due to the intricate combinatorics of the 
interaction vertices in matrix and tensor field theories (TFT),
the analogue of equation \eqref{SDEscalar} 
turns out to be more complicated. As a matter of fact, rank-$D$ tensor models have a propagator  
composed of $D$ parallel lines (nevertheless, denoted by 
$ \subircm{\tikz[scale=.5]{ \draw[dashed] (2,0) -- (.5,0) ;}} $) each of which transmits   
momentum independently from the others, known as \textit{coloring} (see Sec. \ref{Strategy} or \cite[Fig. 2]{fullward}). 
In particular, a quartic interaction vertex
involves a choice of which of those colors are transmitted 
upwards, which downwards, which forwards in the small blob, $V$, of the sunset-term: 
 \begin{align} \label{SDEsNaive}
 \raisebox{-.435\height} 
{\tikz[scale=.87277]{
  \draw[ dashed ] (-2,0) --(2,0 );
  \draw[ dashed] (0,0) circle (1);
 \draw[black,fill=lightgray] (1,0) circle (.5);
\node at (1,0) { $?$};
  \draw[gray,fill=white] (-1,0) circle (.25);
  \draw[fill=gray] (-1.26,0) circle (.05);
    \draw[fill=gray] (-.71,0) circle (.05);
      \draw[fill=gray] (-1+.05,.27) circle (.05);
        \draw[fill=gray] (-1+.05,-.27) circle (.05);
\node at (-1,0) {\scriptsize {$V$}};
 }}\,\,.\qquad  % (V \in \{ \vcCT \}_{c=1,2,3} \cup \{ \tetramin \}  ).  \,\,\,\,\, \nonumber 
 \end{align}
 For $D=3$ and the $2$-point function, notice that 
 for the vertex $V$ of the sunset-diagram, the following might happen:
 \begin{itemize}
\item  $V$
 can be any of $ \{ \vcCT \}_{c=1,2,3}$ or $ V= \tetrach$ 
 for real or $\orth{N}$ tensor models initiated by Carrozza and Tanas\u a \cite{On},
 and used by Klebanov and Tarnopolsky \cite{KlebanovTarnopolsky} in the context 
 of the SYK-like tensor models  
\item or $V= \vc$ for $c=1,2,3$ for complex or $\uni{N}$ tensor models. The black-white 
bipartiteness reflects the presence of both the tensor field and its conjugate. Rather these $\uni{N}$-invariant theories are 
the TFT we shall deal here with.
 \end{itemize} 
 Higher-point functions follow an even more complicated schema\footnote{A simplification is 
 the ``melonic approximation'' \cite{us}}. 
 Either way, accordingly, the Green's functions need further specification
and, in fact,
the classification of the correlation functions 
for higher-rank theories is the following:  
\begin{itemize}
 \item for real matrix theories, there are 
 as many connected
 $k$-point functions \cite{GW12} 
 as integer partitions of $k$. Then, there are 
 three $3$-point functions, five $4$-point functions, seven $5$-point functions, and so on.
\item for complex tensor models  
the connected $k$-point functions are classified
by (possibly disconnected) $D$-colored\footnote{This
is a common abbreviation in the tensor model jargon,
for ``vertex-bipartite regularly edge-$D$-colored graphs''.} graphs in $k$ vertices \cite{fullward}. 
In particular $k=2\ell$ should be even. Each edge in these graphs is of certain color $a$, and 
 this enforces momentum-transmission\footnote{We call the index
 that is transmitted ``momentum'' because these  models are 
 originated in certain Group Field Theory context, whose 
 Fourier dual has the structure of a TFT; 
 if one interpret the group manifolds as direct space,
 then the indices are the momenta.} of this very color.
 Therefore, the Feynman graph structures with four legs 
 can encode momentum-transmission according to 
 $\vuno$, $\vdos$, $\vtres$ or (as pictured in Fig. \ref{fig:nia}) $\meloncito \sqcup \meloncito$ .
 \end{itemize}
% The reason behind the enriched structure of the 
% correlation functions can be seen, combinatorially, as follows: 
%  Tensor models postulate non-local interactions. 
%  For instance, in rank $D=3$, a point-like interaction 
%  can be replaced by any of the enriched quartic vertices $\vuno$, $\vdos$ or $\vtres$.  

% The topological (geometrical) interpretation of this results will be explained
% towards the end of this paper. 
 
In particular, in order to obtain the analogue of eq. \eqref{SDEscalar} 
in a $D$-dimensional QFT-context, say, for 
quartic tensor field theories of rank-$D$, one needs to specify which of the four $4$-point function 
we are inserting into the 2-point function.  The aim of this paper is to explain how to 
achieve this and to arrive at analytic SDE for every (connected)
correlation function. The methods exposed here are
based on \cite{fullward} and \cite{SDE}.
 
\noindent

% \begin{itemize}
% 
% \item \textit{upload .tex files to TikZ-site}
% 
% \item recall that colored $\to$ random or complex tensor models
% 
% \item tensor field theory (explain, not about combinatorics)
% 
% \item graph calculus as generally useful tool 
%  in colored tensor models
%  \item 
% \end{itemize}

% 
% For those models, the functional integration of 
% the partition function yields 
% 
% \[
% a
% \]

\noindent
\textit{Remark.} We do not use Einstein's implicit summation notation.
\section{The strategy} \label{Strategy}

The idea of the utilization of a matrix Ward identity (based on
the Ward identity \cite{DGMR}) in order to 
derive the SDE of matrix models is due to Grosse and Wulkenhaar \cite{GW12,GW09}.  

\subsection{
Complex tensor models}\label{CTM}
Complex tensor models, colored tensor models and random tensor models
\cite{GurauRyan} study fields $\phi_{\xb}\in \C$
($\xb\in I_1\times\ldots\times I_D \subset \Z^D$) whose indices\footnote{We 
think of the large-$N$ limit, so we write, instead of $I_a$, 
directly $\Z$.} transform 
independently under elements of $\uni{N}$
of the product group $H_D=\uni{N}^D$. This means that 
\begin{align*}
\phi_{x_1\ldots x_D}& \mapsto \phi'_{x_1\ldots x_D}= \sum_{y_a}[  W_a]_{x_ay_a} \phi_{x_1\ldots y_a\ldots x_D}\,, \quad 
\bar\phi_{x_1\ldots x_D}  \mapsto \bar\phi'_{x_1\ldots x_D}= \sum_{y_a}[ \overline W_a]_{x_ay_a} \bar\phi_{x_1\ldots y_a\ldots x_D} \,, 
\end{align*}
for each $W_a$ in the $a$-th factor $ \uni{N}$ of $H_D$, for any $a=1,\ldots,D$.
Each of the factors (and of the location of the tensor indices) is referred to as a \textit{color}\footnote{For historical reasons \cite{diFrancesco_rect}.}.
Interactions of this kind of theories are $H_D$-invariants.
We restrict to models for which any (graph)-vertex
lies on a subgraph of the type 
 \raisebox{-.5\height}{
\includegraphics[width=1.8cm]{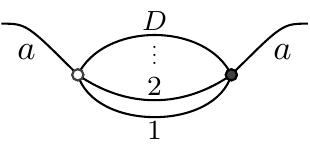}}
for certain color $a$. For $D=3$, this constrains the interactions of
 models to the list\vspace{-.4cm}
  \begin{equation} \label{allowedbubbles}
  \va,
   ~\logoB{6}{Q}{3.82}{23}\,,~\iconoB{8}{Q}{4.7}{23}\,,~\iconoB{10}{Q}{5.8}{23}
\,,~\iconoB{12}{Q}{6.9}{23}\,,\ldots (\mbox{for any color $a$}). % \mbox{(for rank-$3$ models).}
  \end{equation}
Other type of interactions need another methods.
  The origin of this restriction is technical and 
  will be explained in Section \ref{sec:usefulWTI}. 
  Here, we treat models with pillow-like interactions,
  but otherwise without any restriction in their rank. 
\textit{Pillows} are  melonic\footnote{That is, with vanishing Gur\u au-degree \cite{GurauRyan},
but this is concept is not essential here because the present results 
entail no $1/N$-truncation.} graphs 
of four vertices. These are 
$ \{
\vuno,\vdos,\vtres\}, \mbox{  for rank-$3$ models, }\,
 \big\{ \Vuno, \Vdos,\Vtres,\Vcuatro \big\} 
$  for rank-$4$ models, etc. When the rank is clear,
we denote by $V_a$ the
pillow with preferred color $a$ (e.g. $\vuno= V_1$). 
\subsection{The usefulness of the Ward-Takahashi Identity}
\label{sec:usefulWTI}

We consider the quartic tensor model with interaction 
$\Sint = \lambda \sum_{c=1}^D V_c$ with a kinetic Laplacian-like kernel $E$,  
which possibly breaks the $H_D$-symmetry in the quadratic invariant $\Tr_{2}(\bar \phi,\phi) \to S_0=\Tr_{2}(\bar \phi,E\phi)$.
Functional integration of the partition function
yields 
      \[
      Z[J,\bar J; E]   = C \exp \big( 
      -\lambda \textstyle\sum\textstyle_{\mathit a=1}^3 
     V_a \big)\bigg|_{ (\phi,\bar\phi ) \to (\fder{}{\bar J},\fder{}{J}) }   
      Z^{\mtr{free}}[J,\bar J; E]\,.
      \]
Any analytic Schwinger-Dyson equation begins by 
deriving with respect to the sources. By deriving with respect to $\bar J_{\mathbf s}$
we get \cite{fullward,SDE}
\begin{align}
\fder{\log Z [J,\bar J; E]}{\bJ_\mathbf{s}} =
 \frac{1}{E_{\sb}} 
 \bigg\{ J_\sb - \frac{1}{Z[J,\bar J; E]} \bigg(\dervpar{\Sint(\phi,\bar\phi)}{\bar\phi_\sb}\bigg)\bigg|_{ (\phi,\bar\phi ) \to (\fder{}{\bar J},\fder{}{J}) }   
Z[J,\bar J; E] \bigg\} \, .
\end{align}
By assumption (Sec. \ref{CTM}), the term in round parenthesis contains, after evaluation of the sources,
 the subgraph \raisebox{-.5\height}{
\includegraphics[width=1.3cm]{subgraph_without_momenta.pdf}},
and thus a derivative of the form  
  \begin{align}  \label{doubleder}
  \mathcal L_{m_an_a} \equiv
 \sum\limits_{\mathbf p_{\hat a}\in\Z}   \bigg[
  \fder{^2 Z[J,\bar J; E]}{J_{{p_1\ldots  p_{a-1}m_ap_{a+1} \ldots p_D}} 
  \delta \bar J_{{p_1\ldots  p_{a-1}n_ap_{a+1} \ldots p_D}} }\bigg]  
\end{align} 
       acted on by more derivatives.
%        Let $\mathcal I_{m_an_a;\phat}$ with 
           This term resembles 
       the LHS appearing in the Ward-Takahashi Identity (WTI) 
  \begin{align}  \label{eqDeltas}
 \sum\limits_{\mathbf p_{\hat a}\in\Z}   \bigg[
  \fder{^2 Z[J,\bar J; E]}{J_{{p_1\ldots  p_{a-1}m_ap_{a+1} \ldots p_D}} 
  \delta \bar J_{{p_1\ldots  p_{a-1}n_ap_{a+1} \ldots p_D}} } \times E_{m_an_a; \mathbf p_{\hat a}}  \bigg]
    =  D_{J,\bar J}  \,Z[J,\bar J; E]  
\end{align} obtained by Ousmane-Samary \cite{DineWard}
       from the $\uni{N}$-invariance (this group being
the $a$-th factor of $H_D$) of the path 
integral.  Here, 
$E_{m_a n_a; \mathbf p_{\hat a}}\equiv E_{p_1\cdots m_a\cdots p_D}-E_{p_1\cdots n_a\cdots p_D}$,
$\mathbf p_{\hat a}\equiv(p_1,\ldots,p_{a-1} ,p_{a+1},\ldots,p_D)$ 
and $D_{J,\bar J}$ is a first order 
differential operator in the sources. 
It would be useful to reduce the derivatives 
by using this WTI; however, it also
implies the difference of the kernels. 
We restrict, therefore, to models that satisfy that: 
\begin{equation}
\mbox{for any color $a$, }\,\,\,
 E_{m_a n_a; \mathbf q_{\hat a}}=E_{q_1\cdots m_a\cdots q_D}-E_{q_1\cdots n_a\cdots q_D}
 \mbox{ does not depend on $\mathbf q_{\hat a}$.
 }\label{condition} \tag{$\diamondsuit$}
\end{equation}
 We thus write 
only $E_{m_a n_a; \mathbf p_{\hat a}}=:E_{m_a n_a}$ from now on.
The condition \ref{condition}
allows one to get the term $E_{m_a n_a}$ out of the sum
and solve for $\mathcal L_{m_an_a}$. 
We remark that the non-triviality of this task relies on the 
skew-symmetry of the indices of $E_{m_a n_a} $.
This means that we need to find the term 
that is proportional to $\delta_{m_an_a}$ in $  \mathcal L_{m_an_a} $ (see eq. \eqref{doubleder}).
  This is a functional that we denote by $Y\hp{a}_{m_a}[J,\bar J\,]$ (and name, sloppily, $Y$-term). 
  After complete knowledge about this $Y$-term has been obtained we say the WTI is \textit{full}.
The first full WTI was found for $\orth N$ matrix models \cite[Sec. 2]{GW12}.

\section{The full Ward-Takahashi identity}

For arbitrary rank-$D$ $\uni N$ tensor models the full WTI reads \cite{fullward}:
\begin{align} \label{ward2} \!\!\!\!\!
%  &\!\!\!\!\! \!\!\!\!\!\!\!\! \!\!\!
 \mathcal L_{m_a n_a}
% \\
=  
\left(\delta_{m_an_a} Y\hp a_{m_a}[J,\bar J]\right) \,\cdot Z[J,\bar J\,;E]+ \sum\limits_{\phat\in\Z^{D-1} }
\frac1{ E_{m_an_a} } 
 \left( 
\bar J_{m_a \phat} 
\fder{}{\bar J}_{ n_a \phat}  
-J_{n_a \phat}\nonumber 
\fder{ }{J}_{m_a \phat}\right)Z[J,\bar J;E]\,.
\end{align}  
In the next subsections, we explain how to define the correlation functions,
and, subsequently, how to obtain the $Y$-term.

\subsection{The expansion of the free energy in boundary graphs }

The (connected) correlation functions of TFTs will be defined as 
derivatives with respect to sources, as in usual QFT. Nevertheless,
the naive Ansatz
\begin{align} \nonumber 
\log Z[J,\bar J] &  \stackrel{?}{=}
\sum_{\mathbf x,\yb}
G\hp{2}(\mathbf x,\mathbf y)
J_{y_1y_2y_3}
\bar J_{x_1x_2x_3}+\sum_{\mathbf w,\mathbf x,\mathbf y,\mathbf z}
G\hp{4} (\mathbf w,\mathbf x,\mathbf y,\mathbf z)
J_{x_1x_2x_3}
\bar J_{w_1w_2w_3}
J_{y_1y_2y_3}
\bar J_{z_1z_2z_3} + \mathcal O(4)\, 
\end{align}
is, due to the color structure, an oversimplification 
that impairs the derivation of analytic Schwinger-Dyson equations for the thus defined $G\hp{2\ell}$-functions.
The right expansion takes into account the transmission of momentum
inside classes of Feynman graphs, and is given by 
\allowdisplaybreaks[0]
\begin{align} 
\log Z[J,\bar J] &=
\sum_{\mathbf x}
G\hp{2}(\mathbf x)
J_{x_1x_2x_3}
\bar J_{x_1x_2x_3}+\sum_{\mathbf x,\yb }
G\hp{4}_{\mtr{disconn.}}(\mathbf x,\yb)
J_{x_1x_2x_3}
\bar J_{x_1x_2x_3}
J_{y_1y_2y_3}
\bar J_{y_1y_2y_3}\\
&
+\nonumber 
\sum_{\mathbf x,\mathbf y}
G\hp{4}_1(\mathbf x,\mathbf y)
J_{y_1y_2y_3}
\bar J_{x_1y_2y_3}
J_{x_1x_2x_3}
\bar J_{y_1x_2x_3} +
\sum_{\mathbf x,\mathbf y}
G\hp 4_2 (\mathbf x,\mathbf y)
J_{y_1y_2y_3}
\bar J_{y_1x_2y_3}
J_{x_1x_2x_3}
\bar J_{x_1y_2x_3} 
\\ & 
+
\sum_{\mathbf x,\mathbf y}
G\hp{4}_3(\mathbf x,\mathbf y)
J_{y_1y_2y_3}
\bar J_{x_1x_2y_3}
J_{x_1x_2x_3}
\bar J_{y_1y_2x_3} + \mathcal O(6) \nonumber
\end{align}
\allowdisplaybreaks[0]
The sub-indices of the 4-point functions $G\hp4$ will be clear 
soon.
This expansion can be conveniently (compactly) organized by, again, colored graphs. 
This would yield an algorithmic derivation of the $Y$-term. Notice the
effect of the the double derivative on the source-term 
\begin{equation}
\sum_{\phat}
\fder{}{J_{ m_a \mathbf p_{\hat a} }}\fder{}{\bar J_{ n_a \mathbf p_{\hat a} }}
\,\,
\sum_{\xb^1, \cdots, \yb, \cdots \xb^\ell}
G\hp{2\ell} (\xb^1, \ldots, \yb, \ldots \xb^\ell)
[J_{\xb^1} \cdots J_{\yb} \cdots J_{\xb^\ell}]
\cdot
[
\bar J_{\cdots} \cdots \bar J_{\ldots y_a \ldots }\cdots \bar J_{\cdots}\,
]\,. \label{hit}
\end{equation}
Namely, from this equation
it is clear that $Y_{m_a}\hp a$ has contributions 
from `hitting' two sources $J$ and $\bar J$ that 
are connected by an $a$-colored edge in the 
boundary graph (in this term, as a matter of fact, $\bar J_{\ldots y_a \ldots }$ and $ J_{\yb}$). In order to understand this, 
we find convenient to recall what happens in the
for matrix field theories. If this is superfluous for the reader,
Sec. \ref{sec:MatrixExpansion} could be skipped.

\subsubsection{The free energy for real matrix models} \label{sec:MatrixExpansion}
As pointed out in the introduction, the correlation functions (the momenta of the free
 energy $\log Z_{\mtr{matrix}}$) of a general real matrix model
 are classified by integer partitions of $k$. If these partitions are indexed by $\alpha \in \{1,\ldots, P(k)\}$),
 the free energy is expanded as
 \begin{align}\label{eq:Wmatrix}
 \log  Z_{\mtr{matrix}}[J] = \sum_{k=1}^\infty \sum_{\alpha=1}^{P(k)} \frac
 1{\sigma(\alpha)}G\hp k_{\alpha} \star \J(\alpha) = \mbox{`sum over triangulation of boundaries (circles)'}.
 \end{align}
 This is shorthand but is not a formal expression. 
 An integer partition $\alpha=(n_1,\ldots,n_{B(\alpha)})\in \Z_{\geq 0}$ of $k$ 
 (i.e. 
 $k=\sum_{r=1}^{B(\alpha)} n_r$ with $n_{B(\alpha)}\neq 0$, 
 and   $n_i< n_j $ if $i<j$) determines $B(\alpha)$ boundaries,
 $n_r$ of which carry $r$ sources attached. Thus,
 $(\J(\alpha))( \mathbf p^1,\ldots,\mathbf p^{B(\alpha)}):=\prod_{r=1}^{B(\alpha)} J_{p_1^rp_2^r} J_{p_2^rp_3^r} 
 \cdots   J_{p_{n_r}^rp_1^r}$
 and $\mathbf p^r=(p^1_1,\ldots,p_{n_r}^r)$. The star, $\star\,$,
 point-wise sums the product $(G_\alpha\hp k\cdot \J(\alpha))$
 is over the arguments $(\mathbf p^1,\ldots,\mathbf p^{B(\alpha)})$. 
 Further, $\sigma(\alpha)$ is a symmetry factor. 
 \par
 
 The topological significance of this expansion is clear:
 Equation \eqref{eq:Wmatrix} is a sum over triangulations of the boundary of the surfaces 
 that the ribbon graphs triangulate. That is,
 $\alpha$ determines for $n_r$ circles a precise `triangulation by 
 $r$ intervals'. Based on this, one can derive the 
 free energy for complex tensor models. The useful 
 concept there is that of a \textit{boundary graph.}

\subsubsection{The free energy of complex tensor models}

The expansion of the free energy 
also has a geometrical meaning. It is an
 expansion over all triangulations of 
 boundaries, but in higher dimensions. For $D=3$, these
 are triangulations of closed, orientable 
 surfaces. In fact, these are triangulable 
 by bipartite $3$-colored graphs and the 
 sum turns out to be over all the boundary 
 graph that are triangulated by a particular model $\Sint$. 
 The boundaries are characterized in \cite{fullward}.  
The general expansion for any rank reads: 
\begin{align} 
 \log Z_{\,\,\mtr{tensors}}[J,\bar J; E]& = \sum\limits_{\ell =1}^\infty
  \sum\limits_{\substack{\B \,\mtr{is\,\,boundary \,of\, } \Sint  \\ 
\# \mtr {Vertices}(\B)=2\ell}}  \frac{1}{|\Autc(\B)|} 
G^{(2\ell)}_{\B} \star \J(\B) \, . 
\end{align}  
The elements of this formula are:  
\begin{itemize}
 \item to each boundary graph $\B$, 
 $\J$ associates a function $\J(\B):  M_{D\times \ell}(\Z) \to \C$ in the sources
 given by \begin{equation}  \label{JB}
(\J(\B))(\Xb) \equiv  \prod _{\alpha =1 }^\ell J_{\xb^\alpha} \bar J_{\yb^\alpha(\Xb)}, \mbox{being $\xb^\alpha$, ($\alpha=1,\ldots, \ell$) the columns of $\Xb\,$,}
\end{equation}
and $\{\yb^\alpha(\Xb)\}_{\alpha}$ is the set of (unordered) momenta 
$\yb^\alpha$ that one gets in the   $\bar J$-sources 
at the external legs of a Feynman graph $\G$ with $\partial\G=\B$
by `injecting momenta the ${\xb^\gamma}$' 
($\gamma=1,\ldots,\ell$) at the external legs 
marked by $J$-sources (see Fig. \ref{fig:nia}).   
We choose the notation $\B_*(\Xb)=(\yb^1,\ldots,\yb^\ell)$ (see \cite{fullward} for
the detailed construction).
For instance\footnote{Recall
that we do not use Einstein's sum convention. },
$ \J(\vuno)(\xb,\zb) = J_{z_1z_2z_3}
\bar J_{x_1z_2z_3}
J_{x_1x_2x_3}
\bar J_{z_1x_2x_3} $,
since $(\vuno)_*(\xb,\zb)=(x_1,z_2,z_3\,,\, z_1,x_2,x_3)$  
\item one then sums the product $G\hp{2\ell}_\B(\Xb)\cdot (\J(\B))(\Xb)$
% with the monomial $(\J(\B))(\Xb)$ defined by \eqref{JB},
over all momenta $\Xb\in M_{D\times \ell}(\Z)$;
the star $\star$ abbreviates this sum 
\item finally, one divides by the order of the automorphism group 
$\Autc(\B)$ of the graph $\B$. 
The automorphism group will be important  in the following section.

\end{itemize}

\begin{table} \centering
\includegraphics[width=1.0\textwidth]{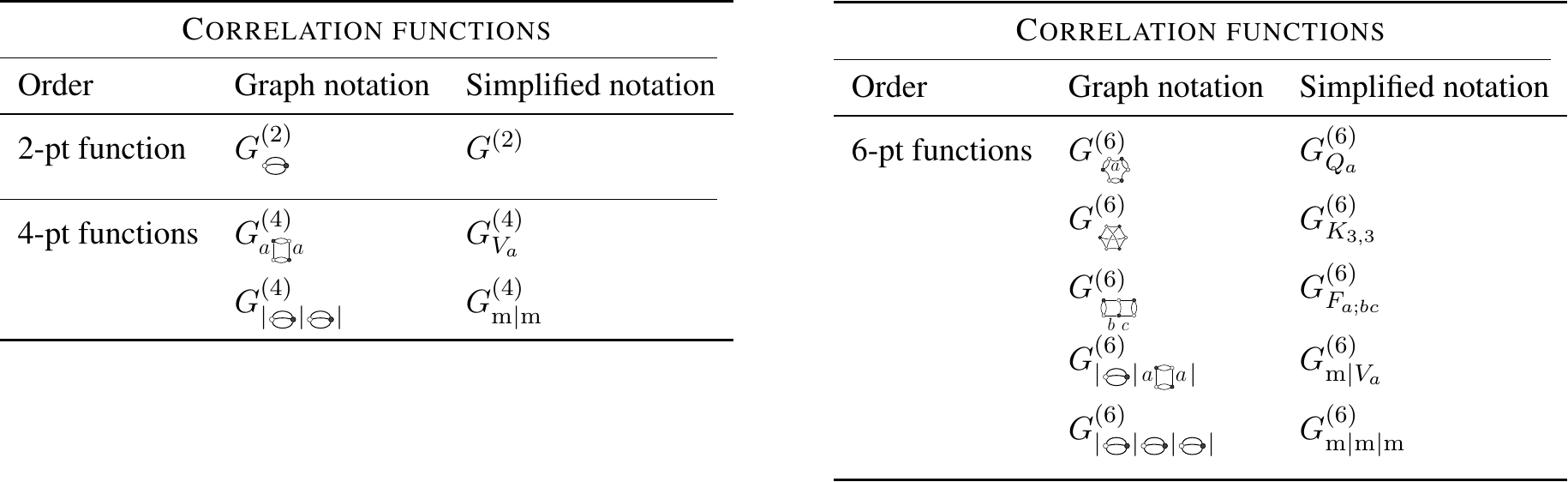}
 \caption{Correlation functions of the $\phi^4_3$-theory 
 until order six \label{table:CR}}
\end{table}

Since from now on we work only with TFTs,
 we omit the subindex `tensors' in the partition function $Z_{\mtr{tensors}}$.
 For rank-$3$ models, the most general expansion is 
 \begin{align}\nonumber
% correct factors
\hspace{-.1cm}
\log Z_{}[J,\bar J\,]&=G\hp 2 \star \J ( \mathrm {m})  + 
\frac1{2!}
G\hp{4}_{\hspace{.2pt}\mathrm{m}|\hspace{.2pt}\mathrm{m}} 
 \star \J (\hspace{.2pt}\mathrm{m}|\hspace{.2pt}\mathrm{m}) + 
\frac1{2} \sum_{c=1}^3 \bigg[
G\hp4_{V_c} \star \J
\big( \, V_c\, \big) 
+ 
% \sum_{c=1}^3
\frac1{3}G\hp 6_{Q_c} 
\star \J(Q_c)
+ 
% \sum_{c=1}^3 
G\hp 6_{F_c} \star \J
(F_c)
\\  
\label{W_expansion_singraficas}
&\,\,\,\,\,\,\,
+
\frac12 G\hp 6 _{\mathrm{m}  \hspace{.2pt}| \hspace{.2pt} V_c} 
\star  \J
(\mathrm{m}  \hspace{.2pt}| \hspace{.2pt} V_c) 
\bigg]
+
\frac1{3} G\hp6 _{K_{3,3}} \star \J (K_{3,3})
+ \frac1{ 3!}
G\hp 6 _{\hspace{.2pt}\mathrm{m}|\hspace{.2pt}\mathrm{m}|\hspace{.2pt}\mathrm{m}\hspace{.2pt}}  \star 
 \J (\hspace{.2pt}\mathrm{m}|\hspace{.2pt}\mathrm{m}|\hspace{.2pt}\mathrm{m}\hspace{.2pt})
% \\   \nonumber
% & \,\, \,\,
% \sum_{c=1}^3 
+ \mathcal O(8) \,.
\end{align}
As shown in \cite{fullward} (relying on \cite{cips}) 
for rank-$D $ models with all\footnote{Actually $D-1$ of the pillows suffice, but that theory is ugly.} 
the pillows $\Sint=\lambda \sum_{c=1}^D V_c$, 
for \textit{any} $D$-colored graph $\B$, even if $\B$ is disconnected, $G\hp{2\ell}_{\B} \equiv\!\!\!\!\!\!\! /\,\,\,\, 0$ holds. 
Table \ref{table:CR} shows the 
transition from the original
source \cite{fullward} notation to the compact one used 
here.

 \subsection{Graph calculus}
The free energy is generated by graphs. Since this is not a
formal expansion, a tool should be developed in order to 
read off the coefficients (functions) of the graphs.
This is the graph calculus \cite{fullward}, which 
consists in deriving functionals $\mathcal A\jj$
with respect to   
$(\J(\B))(\Xb)$, where $\B$ is a boundary graph,
and by momenta $\Xb=(\xb^1,\ldots,\xb^\ell) $
($2\ell = \# $ vertices of $\B$). We restrict $\Xb$
to 
\[ \mathcal{F}_{D,\ell}:=
 \{(\xb^1,\ldots,\xb^\ell) \in M_{D\times \ell} (\Z) \,|\, x^\alpha_{c}\neq x^\nu_c \,\,\mbox{ if }\alpha\neq \nu\,   (\alpha,\nu=1,\ldots, \ell),  \mbox{ for all } 
  c=1,\ldots,D\}\,,
\]
the space of momenta away from the `colored diagonals'. 
One then sets $\partial \mathcal A\jj /\partial \B$ to be the function 
that at $\Xb\in \mathcal{F}_{D,\ell}$ takes the value 
\[\fder{ ^{2\ell}\mathcal A\jj}{\J(\B)(\Xb)}\bigg|_{J=\bar J=0}.\]

An important result is the independence of graphs, meaning that if $\mathcal C$
is another graph 
$\partial \mathcal C/\partial \B$ is non-zero only if the graphs $\B$ and $\mathcal C$
are isomorphic. If that is the case, the derivative $\partial \mathcal C/\partial \B$
is found to be a group action by $\Autc(\B)$. Concretely, 
if $\mathbf C=(\mathbf{c}^1,\ldots,\mathbf{c}^ k)\in
M_{D\times k}(\Z)$ and $\Xb\in \mathcal{F}_{D,\ell}$,
\begin{align*}
\frac{\partial   \mathcal C( \mathbf{C})  }
{\partial\mathcal B(\Xb)}=
\begin{cases} \delta_{\ell k} \cdot  \Big(
 \displaystyle\sum\limits_{\hat \sigma\in\Autc(\mathcal B)}
\delta^{ \mathbf{ c}^{\sigma (1)} ,\ldots,\mathbf{c}^ {\sigma(k)}}
      _{\xb^1,\xb^2, \ldots,\xb^ \ell}\Big)
  & \qquad \mbox{if } \,\,\, \B \cong \mathcal C\, , 
 \\
  \,\,\,\quad 0 & \qquad \mbox{if } \,\, \mathcal B \ncong \mathcal C\,.  
 \end{cases}
\end{align*}
where\footnote{Also, the delta $\delta_{\ell k}$ is somehow redundant (as it is a consequence
of having isomorphic graphs).} $\hat \sigma \in \Autc(\B)$ denotes the lift of a permutation 
$\sigma \in \mtr{Sym}(\ell)$ to the\footnote{The 
colored automorphisms are rigid enough to be specified by
only a permutation of the white (or black) vertices \cite{fullward}. } corresponding element of the 
automorphism group $\Autc(\B)$. Now we are in position to define the correlation functions
 by
 \begin{equation}
 G_\B\hp{2\ell} \equiv \dervpar{\log Z[J,\bar J\, ;E]}{\B}\,\,.
\label{defCR}
 \end{equation}

\begin{example}[Meaning of $\partial /\partial\hspace{1pt} \kthreea$]

If $\A \jj$ is a functional and $\Xb = (\xb^1,\xb^2,\xb^3) \in \mtc F_{3,3}$, one has
\[
\dervpar{\A \jj }{\hspace{1pt}\kthreea}(\Xb)=
\fder{^6 \A \jj}{  J_{\xb^1} \delta J_{\xb^2}\delta J_{\xb^3} 
\delta \bar J_{\yb^1}
\delta \bar J_{\yb^2} \delta \bar J_{\yb^3}} \bigg|_{J=0=\bar J}=
\fder{^6 \A \jj}{  J_{\xb^1} \delta J_{\xb^2}\delta J_{\xb^3} 
\delta \bar J_{x^1_1x^2_2x^3_3}
\delta \bar J_{x^2_1x^3_2x^1_3} \delta \bar J_{x^3_1x^1_2x^2_3}} \bigg|_{J=0=\bar J}\,\,,\]
 since \[ (\hspace{1pt}\kthreegg)_*\{\xb^1,\xb^2,\xb^3\}=  \{\yb^1,\yb^2,\yb^3 \} = \bigg\{\bigg[
 \begin{smallmatrix} x^1_1 \\  x_2^2 \\x_3^3  \end{smallmatrix}\bigg],\bigg[ 
   \begin{smallmatrix} x^2_1 \\ x_2^3 \\x_3^1\end{smallmatrix}\bigg],     \bigg[ \begin{smallmatrix} 
x^3_1 \\ x_2^1 \\ x_3^2 \end{smallmatrix} \bigg] \bigg\}\,.\]
The effect of the operator $\partial /\partial\hspace{1pt} \kthreea$
acting on the free energy is to generate a boundary-torus, since $\kthreea$ is 
a graph triangulating $\T^2$. 
The double action of (say) $\partial^2 /\partial\hspace{1pt} \kthreea \hspace{1pt}\partial\hspace{1pt} \meloncito$ on $\log Z$ 
---denoted by $G\hp 8_{\meloncitito \hspace{1pt}|\hspace{1pt} \kthreea}$
according to eq. \eqref{defCR} --- is to select, from among all the spaces generated by 
the tensor model in question, only the bordisms from the sphere to
the torus, that respect the particular triangulation given by $\meloncito$ and $\kthreea $.
\end{example}
\subsection{The Y-term}

From the given expansion of the free energy 
one can derive the $Y$-term.
In order to do so, we need to introduce the 
functions $\Delta_{m_a,r} G_{\B}\hp{2\ell}$.
These are the coefficients   in eq. \eqref{hit},
after hitting the $r$-th white vertex $\xb^r$
and the vertex $\yb^{\mu(r,a)}$ connected to it by a 
$a$-colored vertex in the boundary graph. To wit,
when the two derivatives act on $J_{\xb^r}$ and $ \bar J_{\ldots x^r_a \ldots }$ in 
\begin{equation}
\sum_{\phat}
\fder{}{J_{ m_a \mathbf p_{\hat a} }}\fder{}{\bar J_{ n_a \mathbf p_{\hat a} }}
\,\,
\sum_{\xb^1, \cdots, \xb^r, \cdots \xb^\ell}
G\hp{2\ell}_{\B} (\xb^1, \ldots, \xb^r, \ldots \xb^\ell)
[J_{\xb^1} \cdots J_{\xb^r} \cdots J_{\xb^\ell}]
\cdot
[
\bar J_{\cdots} \cdots \bar J_{\ldots x^r_a \ldots }\cdots \bar J_{\cdots}\,
]\, , \label{hit2}
\end{equation}
two vertices of the boundary graph are removed. 
The surviving sources have then the form  \[(\J(\B\ominus e^r_a))(\xb^1,\ldots,\widehat{\xb}^r,\ldots, \xb^\ell)\] 
for certain residual graph denoted\footnote{See \cite{fullward} for a deeper
discussion and a more explicit definition. To explain the notation, 
an example would be useful: for $ \mathcal{E}_a = \icono{6}{Eabc}{3.5}{25}$ one has
$\mathcal{E}_a\ominus e^1_a= \vch$, $\mathcal{E}_a\ominus e^2_a= \meloncito \sqcup 
\meloncito$ and $\mathcal{E}_a\ominus e^3_a= \vbh$. Also,
in \cite{fullward} the explicit formula for $\Delta_{m_a,r} G_\B\hp{2\ell}$ 
is given, instead of the rather abstract definition given here.} by $\B\ominus e^r_a$. 
Its coefficient is the function $\delta_{m_an_a}\Delta_{m_a,r} G_\B\hp{2\ell}$,
by definition of $\Delta_{m_a,r}$. Notice that $\Delta_{m_a,r} G_\B\hp{2\ell}$ has $\ell-1$
arguments in $\Z^D$. In this notation, the explicit $Y$-term consequently reads \cite{fullward}: 
\allowdisplaybreaks[1]
\begin{align}
\label{W_expansion_congraficasY}
% correct factors
\hspace{-1cm}
Y\hp a_{m_a} [J,\bar J]&=
   \sum\limits_{q_c,q_b}
G\hp 2 (m_a,q_c,q_b)  
+ \frac12
\sum\limits_{r=1}^2\big( \Delta_{m_a,r}
% ^{{|\raisebox{-.4\height}{\includegraphics[width=.42cm]{gfx_W/icono_melon.pdf}}|
% \raisebox{-.4\height}{\includegraphics[width=.42cm]{gfx_W/icono_melon.pdf}}|}}
G\hp 4 _{\mathrm{m}\hspace{.2pt}|\hspace{.2pt}\mathrm{m}}+\sum_{i=1}^3
\Delta_{m_a,r}
% ^{{\raisebox{-.4\height}{\includegraphics[width=.6cm]{gfx_W/logo_vone.pdf}}}}
G\hp4_i
\big)
\star \J  \big(\mathrm{m}\big)  \nonumber  
\\  
&\,\, + \frac1{3}\sum\limits_{r=1}^3\sum\limits_{i=1}^3
\big(\Delta_{m_a,r}
%^{\raisebox{-.4\height}{\includegraphics[width=.6cm]{gfx_W/logo_resto_6.pdf}}}
G\hp 6_{Q_i}\big)
\star
\J  \big( V_i\, \big) 
 + \frac13 \sum_{r=1}^3 
 (\Delta_{m_a,r}G\hp6_{K_{3,3}})
\star\J\big(V_a\big) 
\nonumber
\\
& \,\,
+\sum\limits_{c\neq a}  \Big\{
\big(\Delta_{m_a,1} G\hp 6_{F_{c;ba}}
\big)
\star\J\big(V_b\big) 
+\big(\Delta_{m_a,2}  G\hp 6_{F_{c;ba}}
\big) 
\star \J
\big(V_b\big)
\\  
\nonumber
&\,\, 
+\big(
\Delta_{m_a,3}  G\hp 6_{F_{c;ba}}
\big)
\star\J
\big(V_a\big)
\Big\}+ \big(
\Delta_{m_a,1}
G\hp 6_{F_{a;bc}}
\big)
\star\J
\big( 
V_c
\big)
+\big(
\Delta_{m_a,2}
G\hp 6_{F_{a;bc}}
\big)
\star\J
\big(\vspace{.3pt}\mathrm{m}\vspace{.3pt}|\vspace{.3pt}\mathrm{m}\vspace{.3pt}
\big) 
 \\
& \,\,
+\big(
\Delta_{m_a,3}
G\hp 6_{F_{a;bc}}
\big)
\star\J
\big( \nonumber 
V_b
\big)
+\frac1{ 3!}
\sum\limits_{r=1}^3\big(\Delta_{m_a,r}
G\hp 6_{\mathrm{m}\hspace{.2pt}|\hspace{.2pt}\mathrm{m}|\hspace{.2pt}\mathrm{m}}\big)
\star\J
\big(\vspace{.3pt}\mathrm{m}\vspace{.3pt}|\vspace{.3pt}\mathrm{m}\vspace{.3pt}\big) 
\\  
\nonumber
&\,\, 
+\nonumber \sum_{i=1,2,3}\Big\{
\frac12 \big(\Delta_{m_a,1}G\hp 6 _{\hspace{.2pt}\mathrm{m}|\hspace{.2pt} V_i }\big)
\star \J
\big(V_i
\big) 
+ \frac12\sum\limits_{r=2,3}\big(\Delta_{m_a,r}
G\hp 6 _{\hspace{.2pt}\mathrm{m}|\hspace{.2pt} V_i }\big)
\star \J
\big(\vspace{.3pt}\mathrm{m}\vspace{.3pt}|\vspace{.3pt}\mathrm{m}\vspace{.3pt}\big) \Big\}
+ \mathcal O(6)\,.
% \\
% & \,\,+ \mathcal O(6)\,. 
% \nonumber 
\end{align}
the notation $F_{a;bc}$
means the graph $F_a=\icono{6}{Eabcs}{3}{45}$ ($b\neq a\neq c\neq b$)
with a left-to-right ordering of the white vertices.
Also, the graph-subindex notation (i.e. switching back to 
the left columns of Table \ref{table:CR})
might be helpful in order to understand how this expression 
was computed.
  \\
% For the next section, it is important to write 
% 
% \begin{equation}
% Y\hp a_{m_a}\jj = \sum_
% \end{equation}

\section{The tower of Schwinger-Dyson equations (connected boundary)}
 
With the $Y$-term known, it is clear that one can express it as a sum over graphs in the form 
$Y\hp a_{m_a}\jj= \sum_{\mathcal D}  \mathfrak f_{\mathcal D,m_a}\hp{a} \star \J(\mathcal D)$.
The graph calculus allows to compute these $\mathfrak{f}$-functions. 
In order to state the SDE tower, we only need a last graph operation, the swap $\varsigma_a$. \par
Let $\varsigma_a(\B;v,w)$ swap of the $a$-colored edges at two 
black vertices $v,w$ of a colored graph $\B$.  
Examples of this operation are\footnote{Notice that in $\varsigma_a(\kthree;v,w) $ 
there is no dependence on the choice of the vertices $v$ and $w$, due to the 
symmetries of the graph $\kthree$.} 
\[\,\,\varsigma_a(\kthreegg;v,w) = \logos{6}{Eabc}{4.0}{25} % \mbox{ (independent of $v$ and $u$) and }
\qquad\mbox{ and } \qquad\,\,\varsigma_b( \logos{6}{Eabc}{4.0}{25};\mbox{\small{left}},
\mbox{\small{down}})=\melon \sqcup \logos{4}{Vc}{3.0}{33}\,\,.\] 
In order to derive the SDE for $G_{\B}\hp{2\ell}$ 
one has to choose a black vertex of $\B$; thus, in particular,
if $\B$ has no automorphisms (as e.g. $F_c$ in rank 3)
there are $\ell$ independent SDE for $G\hp{2\ell}_\B$.
Derivatives with respect to the graphs $\varsigma_a(\B ;i,n)\equiv\varsigma_a(\B ;\,\yb^i,\,\yb^n) $, 
$n\neq i$,  appear in the SDE. 
\\

Let $\B$ be a connected boundary graph of the quartic rank-$D$ model with pillow interactions, $\Sint=\lambda \sum_{c=1}^D V_c$.
Let $\B$ have $2\ell $ vertices.  The $(2\ell)$-point Schwinger-Dyson
equations corresponding to $\B$ are  \cite[Thm. 3.1]{SDE}
% \leqnomode
\allowdisplaybreaks[0]
\begin{align}
& \hspace{-1cm}\bigg(1+\frac{2\lambda}{E_{\sb}} \suml_{a=1}^D \suml_{\mathbf q_{\hat a} }\GDmelon (s_a,\mathbf q_{\hat a})\bigg)
G_\B\hp{2\ell}(\Xb) \\
& =\frac{\delta_{\ell,1} }{E_\mathbf{s}}+\frac{(-2\lambda)}{E_{\sb}} \suml_{a=1}^D \Bigg\{ 
\suml_{\hat \sigma \in \Autc(\B)} \sigma^* \mathfrak{f}\hp{a}_{\B,s_a}(\Xb) \nonumber
 \label{S_eq:SDEs}
%   & \qquad\qquad \qquad 
%  \,\,\,\,\,
 \\
 &\qquad \quad   +
 \suml_{n\neq i }  \frac{Z_0\inv}{  E_{y^n_a,s_a}} \bigg[\dervpar{Z[J,J]}{\varsigma_a(\B ;i,n) }  (\Xb)
- \dervpar{Z[J,J]}{\varsigma_a(\B ; i,n) } (\Xb|_{s_a \to y_a^{n}})\bigg] \nonumber
 \\
& \nonumber 
\,\,\,\,  \qquad\qquad\qquad - \suml_{b_a}\frac{1}{E_{s_a,b_a}} \big[ G_\B\hp{2k}(\Xb) - 
G_{\B}\hp{2k}(\Xb|_{s_a \to b_a}) \big] \Bigg\}   
\end{align} 
with  $\sb=\yb^i$ picked from  
$\B_*(\Xb)=(\yb^1,\dots,\yb^\ell)$, $1\leq i \leq \ell$ .  Here $\sigma\in\Sym(\ell)$
acts by permuting the arguments of $\mathfrak{f}\hp a_{\B,s_a}$. 
More explicit formul\ae $ $  are given in \cite{SDE}  for 
ranks three, four and five.
\section{Conclusions}

The tools leading to the tower of SDE for arbitrary-rank TFTs
with pillow interactions have been exposed. The 
kernel in the kinetic term should satisfy the mild condition \ref{condition}.
The scope of this method is boarder than only 
pillow interactions (e.g. for rank-$3$ TFTs, 
the list \ref{allowedbubbles}). The obtained equations
are for (connected) correlation functions with connected boundary graph.
The general result for arbitrary, disconnected graphs is work in progress,
as is the extension of the present
methods to fermionic fields and to $\orth N$ TFTs \cite{On}, aiming at SYK-like tensor models.
\\

\bibliographystyle{JHEP}
%    \setstretch{1}
%   \setlength\bibitemsep{2pt} 
 
\providecommand{\href}[2]{#2}\begingroup\raggedright\endgroup 
% \end{thebibliography}

%   \bibliography{BibliographyPostdoc}
%  \bibitem{...} ....
% \end{thebibliography}

\end{document}